# Covalent 2D $Cr_2Te_3$ ferromagnet

Mengying Bian[†], Aleksandr N. Kamenskii[†], Mengjiao Han[†], Wenjie Li, Sichen Wei, Xuezeng Tian, David B. Eason, Fan Sun, Keke He, Haolei Hui, Fei Yao, Renat Sabirianov, Jonathan P. Bird, Chunlei Yang, Jianwei Miao, Junhao Lin*, Scott A. Crooker*, Yanglong Hou*, Hao Zeng*

Dr. M. Bian, Prof. Y. Hou

Beijing Key Laboratory for Magnetoelectric Materials and Devices and Beijing Innovation Center for Engineering Science and Advanced Technology, Department of Materials Science and Engineering, Peking University, China

Email: hou@pku.edu.cn

Dr. M. Bian, F. Sun, H. Hui, Prof. H. Zeng

Department of Physics, University at Buffalo, State University of New York, USA

Email: haozeng@buffalo.edu

A. N. Kamenskii, Prof. S. A. Crooker

National High Magnetic Field Laboratory, Los Alamos National Laboratory, USA

Email: crooker@lanl.gov

A. N. Kamenskii

Experimentelle Physik II, Technische Universita¨t Dortmund, Germany

M. Han, Prof. J. Lin

Department of Physics and Shenzhen Key Laboratory for Advanced Quantum

Functional Materials and Devices, Southern University of Science and Technology,

China

Email: linjh@sustech.edu.cn

Prof. W. Li, Prof. C. Yang

Center for Information Photonics and Energy Materials, Shenzhen Institutes of

Advanced Technology, Chinese Academy of Sciences, China

S. Wei, Prof. D. B. Eason, Prof. F. Yao

Department of Materials Design and Innovation, University at Buffalo, The State

University of New York, USA

X. Tian, Prof. J. Miao

Department of Physics & Astronomy and California NanoSystems Institute,

University of California, Los Angeles, USA

K. He, Prof. J. P. Bird

Department of Electrical Engineering, University at Buffalo, the State University of New York, USA

Prof. R. Sabirianov

Department of Physics, University of Nebraska-Omaha, USA

**Keywords**: 2D magnets; half-metallicity; $Cr_2Te_3$; MOKE microscopy; magnetic anisotropy

**Abstract**

To broaden the scope of van der Waals 2D magnets, we report the synthesis and magnetism of covalent 2D magnetic $Cr_2Te_3$ with a thickness down to one-unit-cell. The 2D $Cr_2Te_3$ crystals exhibit robust ferromagnetism with a Curie temperature of 180 K, a large perpendicular anisotropy of $7\times10^5$ J m$^{-3}$, and a high coercivity of ~ 4.6 kG at 20 K. First principles calculations further show a transition from canted to collinear ferromagnetism, a transition from perpendicular to in-plane anisotropy, and emergent half-metallic behavior in atomically-thin $Cr_2Te_3$, suggesting its potential application for injecting carriers with high spin polarization into spintronic devices.

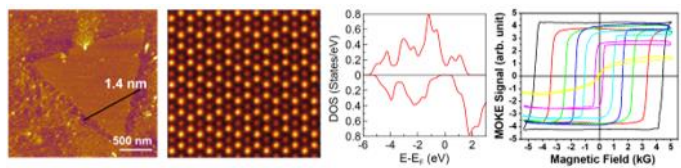

**Impact Statement**

A $T_C$ of 180 K, a perpendicular anisotropy of $7\times10^5$ J m$^{-3}$, and a high coercivity of ~ 4.6 kG were achieved in covalent $Cr_2Te_3$ 2D ferromagnets.

# 1. Introduction

The recent discovery of 2D vdW magnets such as $CrI_3$, $CrGeTe_3$, $Fe_3GeTe_2$, $MnPS_3$ and $FePS_3$, [1, 2, 3], has intensified research in 2D magnetism and spurred interests in searching for 2D magnets with novel properties. 2D magnets have potential technological impact not only because they represent the ultimate scaling limit in data storage and computing devices, but also due to their sensitivity to external perturbations such as electric gating [4, 5]. They can further be integrated into vdW heterostructures, allowing the tuning of magnetic and topological order [6, 7], valley splitting and polarization using proximity effects [8, 9].

The vdW 2D magnets exfoliated from bulk crystals [10] suffer drawbacks such as poorly controlled layer thickness, size and shape in a single flake, low magnetic ordering temperature, and mostly antiferromagnetic interlayer coupling. To address such limitations, it is desirable to explore material systems beyond vdW crystals. To this end, a class of covalent magnetic transition metal chalcogenides exemplified by $Fe_xSe_y$ and $Cr_xTe_y$ are worth exploring [11, 12]. They possess a structure with monolayers of $FeSe_2$ and $CrTe_2$ intercalated by a layer of transition metals with ordered vacancies. The weakened interlayer bonding due to cation vacancies thus offers the possibility to obtain 2D magnets. Previously, we reported a giant coercivity of 4 Tesla and a transition temperature above 300 K in chemically synthesized $Fe_3Se_4$ nanostructures down to a few unit cell thickness [13]. Chemically synthesized $Cr_2Te_3$ nanostructures, thin films grown by molecular beam epitaxy, and 2D layers synthesized by chemical vapor deposition (CVD) have been reported recently [14, 15, 16, 17]. However, systematic theoretical and experimental studies on the magnetic behavior of atomically-thin $Cr_2Te_3$ are so far lacking.

In this work, we report a concerted theoretical and experimental investigation of magnetism in 2D $Cr_2Te_3$ synthesized by CVD down to a single unit-cell thickness. Theoretical calculation predicts ferromagnetic order with an estimated $T_C$ of ~ 200 K and a large perpendicular magnetocrystalline anisotropy of up to $1.1 \times 10^6$ J m$^{-3}$ in the bulk, while making a surprising transition to in-plane anisotropy for atomically-thin $Cr_2Te_3$ from mono- to quad-layers. A collinear ferromagnetic spin arrangement is found for atomically-thin $Cr_2Te_3$, despite spin canting in the bulk. A half-metallic behavior, absent in bulk $Cr_2Te_3$, is predicted to emerge in atomically-thin layers and strengthened by reducing thicknesses. Experimentally, we found a $T_C$ of ~ 180 K, a perpendicular anisotropy of $7 \times 10^5$ J m$^{-3}$ and a coercivity of 4.6 kG at 20 K, from both ensemble magnetic measurements and magneto-optical Kerr effect (MOKE) measurements on single 2D crystals. Our work suggests that expanding beyond vdW materials is promising for discovering 2D magnets with unconventional properties.

## 2. Results and discussion

Atomically-thin $Cr_2Te_3$ crystals were prepared by CVD incorporating a space-confined strategy [18]. In this process, the precursors including Te and $CrCl_3$ powder mixed with NaCl were heated in separate temperature zones of a tube furnace. A schematic of the growth process is shown in **Figure 1**a and details of the growth are provided in the Supporting Information (SI). To obtain 2D layers of covalent $Cr_2Te_3$, the thermodynamic 3D growth needs to be suppressed and thus the growth should be kinetically controlled. The space-confined strategy deploys two substrates stacked together to introduce a local environment of precursor vapor in between (Figure 1b) [18]. Comparing with the conventional technique (Figure 1c), this method can lead to more stable and slower gas flow, and thereby sustain a more steady and suppressed growth rate [19, 20]. The decreased nucleation density is preferred for lateral growth as edge sites are more reactive. Meanwhile, new nucleation on top of 2D layers is discouraged due to weak inter-layer bonding and vapor concentration below the nucleation threshold. Moreover, the vdW mica substrate improves the mobility of the precursor atoms. The combined effects result in atomically-thin flakes as shown in the optical microscope image (Figure 1d). By comparison, using the conventional approach, the density of the crystals obtained is much higher and the thicknesses are also larger (Figure 1e). Figure 1f and 1g show a typical atomic force microscope (AFM) image and the corresponding height profile of a single $Cr_2Te_3$ flake. The thickness is found to be ~ 1.4 nm, which corresponds to a one-unit cell thick $Cr_2Te_3$ crystal consisting of a bilayer $CrTe_2$ and a Cr layer with vacancies in between (**Figure 2**b).

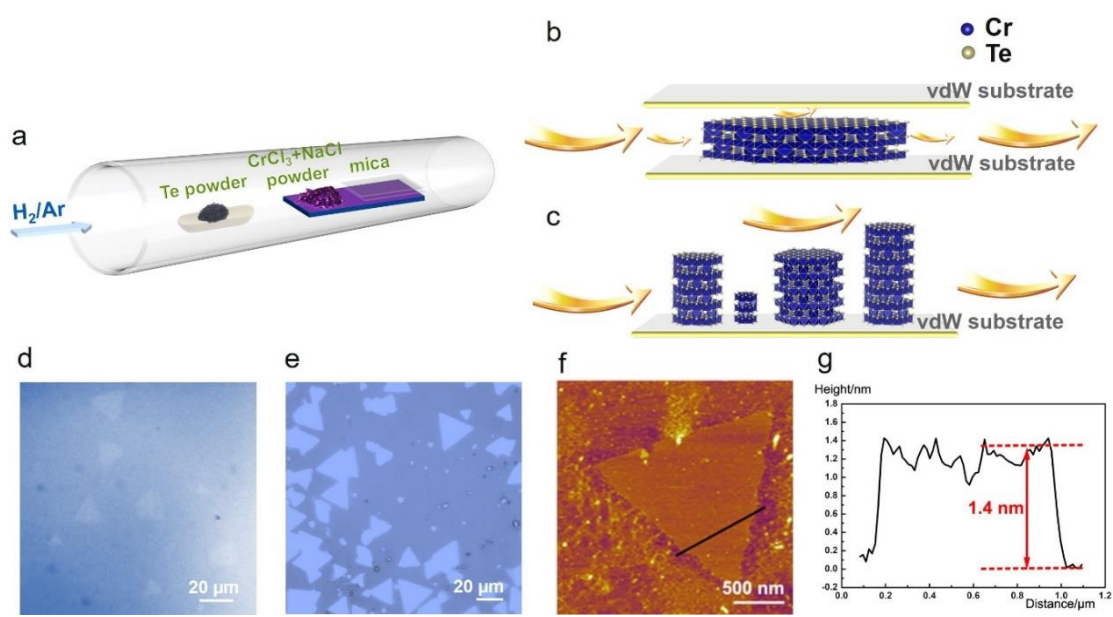

**Figure 1. a,** A schematic diagram of the CVD growth process. Schematic illustrations of the growth processes of $Cr_2Te_3$ using **b,** space-confined and **c,** conventional CVD, respectively. An optical microscope image of $Cr_2Te_3$ grown on mica utilizing **d,** the space-confined CVD and **e,** conventional CVD process. **f,** An AFM image and **g,** The height profile of a single $Cr_2Te_3$ flake. The thickness is ~ 1.4 nm, corresponding to a single unit cell.

Atomically resolved high angle annular dark field scanning transmission electron microscopy (HAADF-STEM) measurements and energy dispersive X-ray spectroscopy (EDS) mapping were carried out to determine the crystal structure and elemental composition of the as-synthesized 2D crystals. $Cr_2Te_3$ possesses a hexagonal structure with a $\bar{P}$-31c (No. 163) space group (a = 6.8 Å, c = 12.1 Å). The Cr atoms classified by different Cr neighbors within the unit cell are labeled as $Cr_I$, $Cr_{II}$ and $Cr_{III}$ (see Figure 2a and 2b). $Cr_I$ is located in between the $CrTe_2$ monolayers with ordered vacancies, whereas $Cr_{II}$ and $Cr_{III}$ are in the $CrTe_2$ layer. $Cr_{II}$ has no Cr neighbor and $Cr_{III}$ has only one $Cr_I$ neighbor along the $c$ axis. $Cr_I$ atoms have two $Cr_{III}$ neighbors along the $c$ axis but no direct Cr neighbor in the *ab* plane. In the top view of the crystal structure (Figure 2a), the octahedral sites form a hexagon with two equilateral triangles connecting the three bottom and three top Te atoms, respectively. Therefore, as manifested in the HAADF-STEM image (Figure 2c) and corresponding atom-by-atom EDS mapping (Figure 2d), the Cr and Te columns can be clearly seen to form a hexagonal pattern with alternating dim (Cr columns) and bright spots (Te columns). This is in accordance with the (100) crystal planes of hexagonal structured $Cr_2Te_3$ (PDF#29-0458). The HAADF-STEM and atomic EDS mapping also synergistically indicate that there is no Te/Cr intermixing, consistent with our atomic $Cr_2Te_3$ model and excluding the existence of anti-site defects. Furthermore, the corresponding Fast Fourier Transform (FFT) pattern (the upper right inset of Figure 2c) reveals a single set of spots with an in-plane three-fold symmetry, which is in agreement with the (120) and (300) lattice periodicity of $Cr_2Te_3$. EDS mapping of Cr and Te of another triangular crystal is shown in Figure S1,. The ratio of Cr and Te is determined as 0.67, consistent with the stoichiometry of $Cr_2Te_3$.

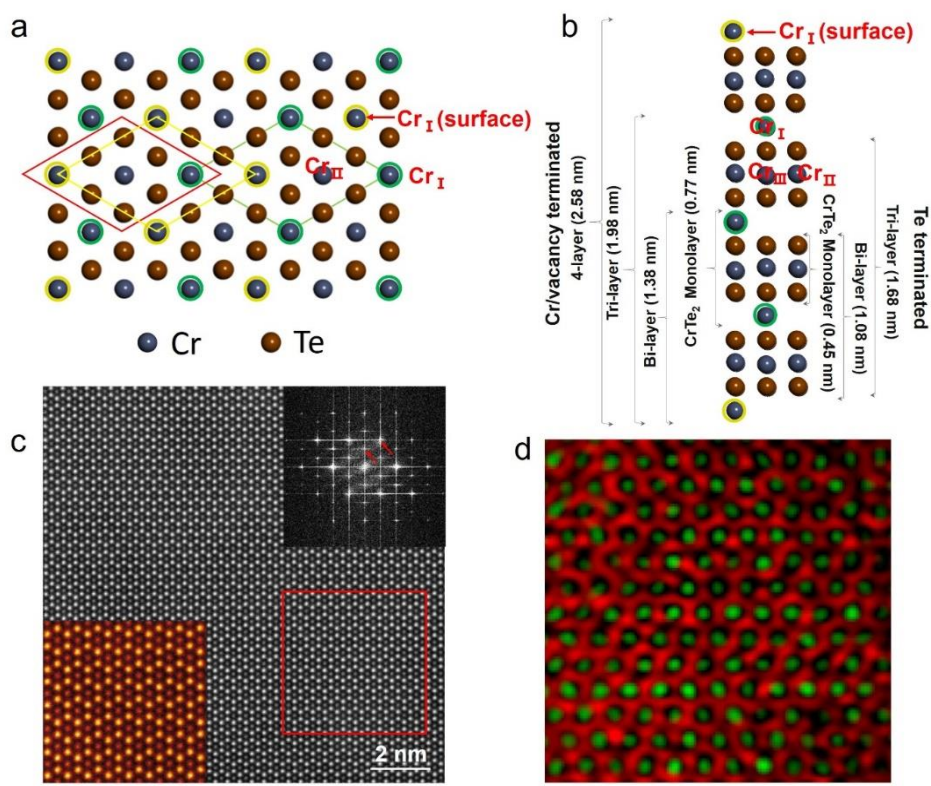

**Figure 2. a,** c-axis projection: unit cell of $Cr_2Te_3$ is shown in the red box; atoms with green halo are $Cr_I$ atoms in the interior Cr/Vacancy layers; atoms with yellow halo are $Cr_I$ atoms on the surface Cr/Vacancy layers. **b,** a-axis projection: braces show the structures used in calculations of single- to quad-layer. The Cr and Te columns are denoted by blue and brown solid columns, respectively. **c,** Atomically resolved HAADF-STEM image of single crystal $Cr_2Te_3$. Upper right inset: the reciprocal lattice obtained from an FFT of the HAADF-STEM image, with the lattice planes of $Cr_2Te_3$ highlighted
9

by red arrows. Lower left inset: an enlarged image of **c**. **d,** EDS elemental mapping of the area shown by the red square in **c**. Green: Cr, Red: Te.

To model atomically-thin $Cr_2Te_3$, we consider thin films with mono-, bi-, tri- and quad- layers of $CrTe_2$ intercalated by a layer of Cr with ordered vacancies (a bi-layer corresponds to a single unit cell in bulk $Cr_2Te_3$). The top and bottom surfaces can be terminated by either Te or Cr atoms. The calculated thicknesses of 1-4 layers terminated by either Te or Cr/vacancy are shown in Figure 2b. Layers terminated by Te atoms experience substantial in-plane contraction compared to bulk $Cr_2Te_3$, and are thus not stable in the 2D limit. However, if these 1-4 layers are terminated with partially filled Cr layers, the surface tension is substantially smaller, and the lattice parameters become closer to those of bulk $Cr_2Te_3$. Therefore, in subsequent calculations, these atomically-thin $Cr_2Te_3$ layers are modeled as being terminated by Cr with ordered vacancies.

From the DFT calculations, magnetic moments of 3.045, 3.126 and 3.096 $\mu_B$ are found for $Cr_I$, $Cr_{II}$, and $Cr_{III}$ respectively in bulk $Cr_2Te_3$. The Te atoms are antiferromagnetically polarized with a moment of -0.195 $\mu_B$. The total magnetization per unit cell (8 Cr and 12 Te atoms) is thus 24.23 $\mu_B$, equivalent to 3.03 $\mu_B$ per Cr assuming a co-linear spin orientation. The calculated magnetization is substantially higher than the experimental value. This can be attributed to spin canting of the Cr atoms in between the $CrTe_2$ layers[21], as discussed later.

The exchange interactions were evaluated using total energy calculations for fully relaxed magnetic configuration at experimental lattice parameters. The pair exchange interactions were mapped onto the Heisenberg model with two nearest neighbors. The Heisenberg model total energy is

$$H = -\sum_{i>j}^{NN} J_{ij}\vec{S_i} \cdot \vec{S_j} \qquad (1),$$

where $J_{ij}$ are pair exchange parameters and $\vec{S_i}$ are unit vectors representing the direction of local magnetic moments. The calculated on-site exchange parameters ($J_0 = -\sum_j^{NN} J_{0j}$) for different Cr sites of 1-4 layers together with that of bulk $Cr_2Te_3$ are given in Table S1. The magnetization and magnetic anisotropy energy (MAE) are shown in Table 1.

**Table 1. Magnetization and MAE of 1-4 layers together with bulk $Cr_2Te_3$.**

|  | 1-L (FM) | 2-L (FM) | 3-L (FM) | 4-L (FM) | Bulk (canted) |
|---|---|---|---|---|---|
| M ($\mu_B$ unit cell$^{-1}$) | 18 | 30 | 42 | 54.15 | 19.2 |
| M ($10^3$ A m$^{-1}$) | 538 | 500 | 488 | 483 | 365 |
| MAE (meV unit cell$^{-1}$) | -8.4 | -9.6 | -4.87 | ~ 0 | 3.5 |
| MAE (MJ m$^{-3}$) | -5.3 | -4.1 | -0.9 | -2.6×10$^{-5}$ | 1.11 |

Note: The unit cell is shown in the rhombic box in Figure 2a. For 1-4 layers, the thickness of the unit cell is the thickness of the layers shown in Figure 2b. For bulk $Cr_2Te_3$, the thickness of the unit cell corresponds to that of a bi-layer without the surface Cr atoms.

The Heisenberg model considering only the nearest neighbor and second nearest neighbor exchange interactions predicts a canted spin structure in bulk $Cr_2Te_3$, as shown schematically in **Figure 3**a. It is found that $J_{12}$ = -3.25 meV, $J_{13}$ = 1.25 meV, $J_{23}$ = 2.92 meV. The competing exchange interactions lead to spin frustration and canting, primarily for $Cr_I$ atoms in the vacancy layer, with much smaller angle for $Cr_{III}$. As can be seen from Figure 3b, canting of $Cr_I$ occurs at relatively large range of $J_{13}/J_{12}$ values. At low $J_{13}$, coupling is antiferromagnetic, which switches to ferromagnetic as $J_{13}$ increases. The black dot represents the canting angle $\theta_1 \sim 81°$ of $Cr_I$ calculated from the exchange parameters obtained using the experimental lattice constant. This leads to a significant reduction of magnetization to 19.2 $\mu_B$/unit cell (2.37 $\mu_B$ per Cr), or $365 \times 10^3$ A m$^{-3}$ (Table 1).

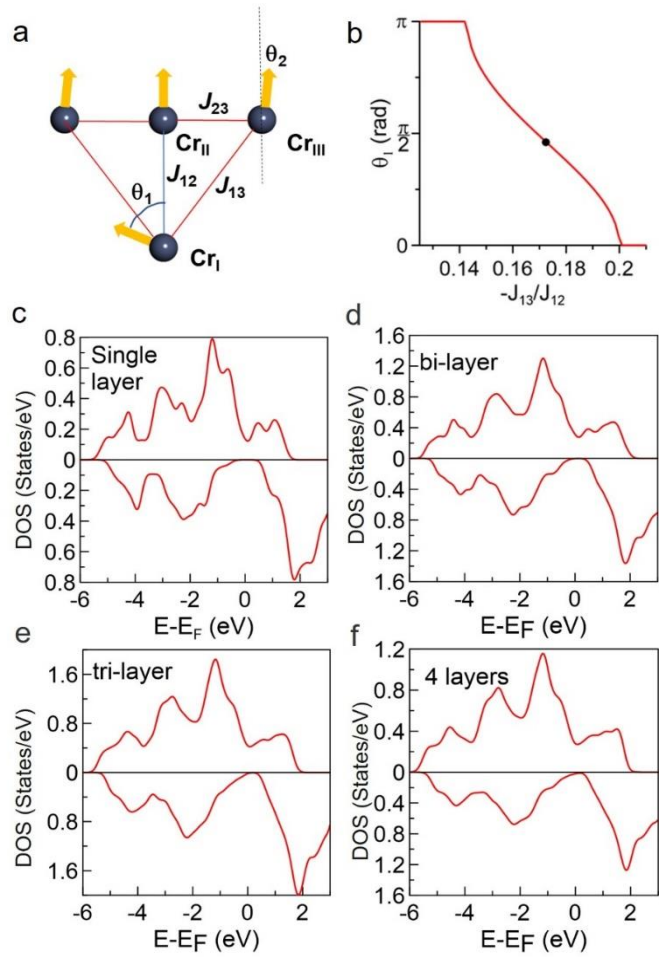

**Figure 3. Competition of spin exchange, spin canting and spin-resolved total density of states (DOS). a,** Schematic representation of exchange coupling in $Cr_2Te_3$. Competing inter-layer exchange parameters $J_{12}$ and $J_{13}$ induce canting angles $\theta_1$ for $Cr_I$ and $\theta_2$ for $Cr_{III}$ atoms. **b,** $\theta_1$ as a function of the ratio of next-nearest to nearest

neighbor exchange parameters $J_{12}/J_{13}$; the black dot represents $\theta_1$ calculated from $J_{12}/J_{13}$ obtained using the experimental lattice constant. Spin-resolved total DOS of **c,** single, **d,** bi-, **e,** tri-layer and **f,** quad-layers $Cr_2Te_3$ calculated for Cr-terminated structures. Majority DOS is shown in the upper panels while minority DOS is shown in the lower panels. $E_F$ is the Fermi energy.

Interestingly, spin canting is suppressed in mono- to quad-layers (i.e. collinear ferromagnetic). More strikingly, the 2D $Cr_2Te_3$ layers exhibit half-metallic behavior; *i.e.* they possess a metallic density of states (DOS) in one spin band and are insulating in the other[22], despite being metallic in bulk $Cr_2Te_3$. As shown in Table 1, the total magnetic moment per unit cell are 18, 30 and 42 $\mu_B$ for the mono-, bi- and tri-layer, respectively. An integer spin moment is a hallmark of half-metallic behavior[23].

As can be seen from the spin-resolved total DOS in Figure 3c-3e, a gap clearly exists in the minority DOS for mono-, bi- and tri- layer $Cr_2Te_3$, confirming their half-metallicity. In the quad-layer, the gap shrinks to near zero (Figure 3f). In bulk $Cr_2Te_3$, such a gap is absent (Figure S3b); however, the minority DOS is extremely low near the Fermi level, suggesting a high spin polarization. Unlike in conventional materials where half-metallic behavior is weakened or destroyed at the surface or interface, the half-metallic behavior in 2D $Cr_2Te_3$ is strengthened by an increase of the gap with decreasing layer thickness [24]. This is due to the lower coordination number in atomically-thin layers, which narrows the band dispersion. The unique 2D half-metallicity in $Cr_2Te_3$ makes it promising as an electrode for magnetic tunnel junctions and spin transistors to realize 100% spin-polarized carrier transport [25].

MAE (Table 1) is strongly perpendicular for bulk $Cr_2Te_3$. Its magnitude is about $1\times10^6$ J m$^{-3}$, of the same order to those of permanent magnets [26]. The large MAE originates from the hexagonal structure and strong spin-orbit coupling due to the presence of heavy element Te. Strikingly the anisotropy turns in-plane (negative) for mono- to quad-layer $Cr_2Te_3$ as shown in Table 1, a behavior that is not understood at present. The MAE is negligibly small for the quad-layer, suggesting that it is close to the critical thickness to a transition from in-plane to perpendicular anisotropy.

The Curie temperature estimated from the on-site exchange parameter as $T_C = \frac{2\langle J_0\rangle}{3k_B}$ is ~306 K for bulk $Cr_2Te_3$, which increases slightly in 2D layers. This is contrary to the behavior of most 2D magnets which show reduced $T_C$. Note that mean-field results typically overestimate $T_C$ by 30-40%. Thus the experimental $T_C$ is expected to be around 200 K for both 2D and bulk $Cr_2Te_3$.

The magnetic hysteresis loops of an ensemble of 2D $Cr_2Te_3$ crystals on mica were measured. **Figure 4**a shows the out-of-plane magnetic hysteresis loops for $Cr_2Te_3$ at temperatures from 10-150 K. A coercivity ($H_C$) of 4.6 kG at 10 K is observed, about 60% smaller than the 1.3 Tesla of bulk films (see Figure S6, SI). The smaller $H_C$ can be explained by lower coordination numbers at sharp corners and edges of 2D crystals, which create local anisotropy smaller than that of the interior and lead to nucleation and propagation of magnetic domains at fields smaller than the anisotropy field. The 10 K loop shows a linear slope persisting to high fields, similar to that of thick films (Figure

S6, SI). This slope is attributed to spin canting discussed in the theory section above. With a canting angle of 81°, contribution of the $Cr_I$ moment to the total magnetization is insignificant at zero field. With increasing field, however, the spins of $Cr_I$ atoms tilt slightly towards the z-axis, progressively increasing the magnetization. To estimate the value of MAE, both in-plane and out-of-plane hysteresis loops were measured at 10 K and displayed in Figure 4b. Clearly the sample shows uniaxial anisotropy with an easy axis perpendicular to the plane, as evidenced by a perpendicular loop that is squarer and a parallel loop with smaller opening and lower remanent magnetization. By extrapolating both hysteresis loops to high fields, the anisotropy field is estimated to be 5.5 T. Combining the saturation magnetization ($M_S$) of $240 \times 10^3$ A $m^{-1}$ measured from a bulk film (see SI), the uniaxial MAE is estimated to be $7 \times 10^5$ J $m^{-3}$. This is slightly smaller than the predicted value, but of the same order to that of hard magnetic CoPt and NdFeB [28]. The appreciable MAE plays a crucial role in stabilizing the long-range magnetic order in 2D $Cr_2Te_3$.

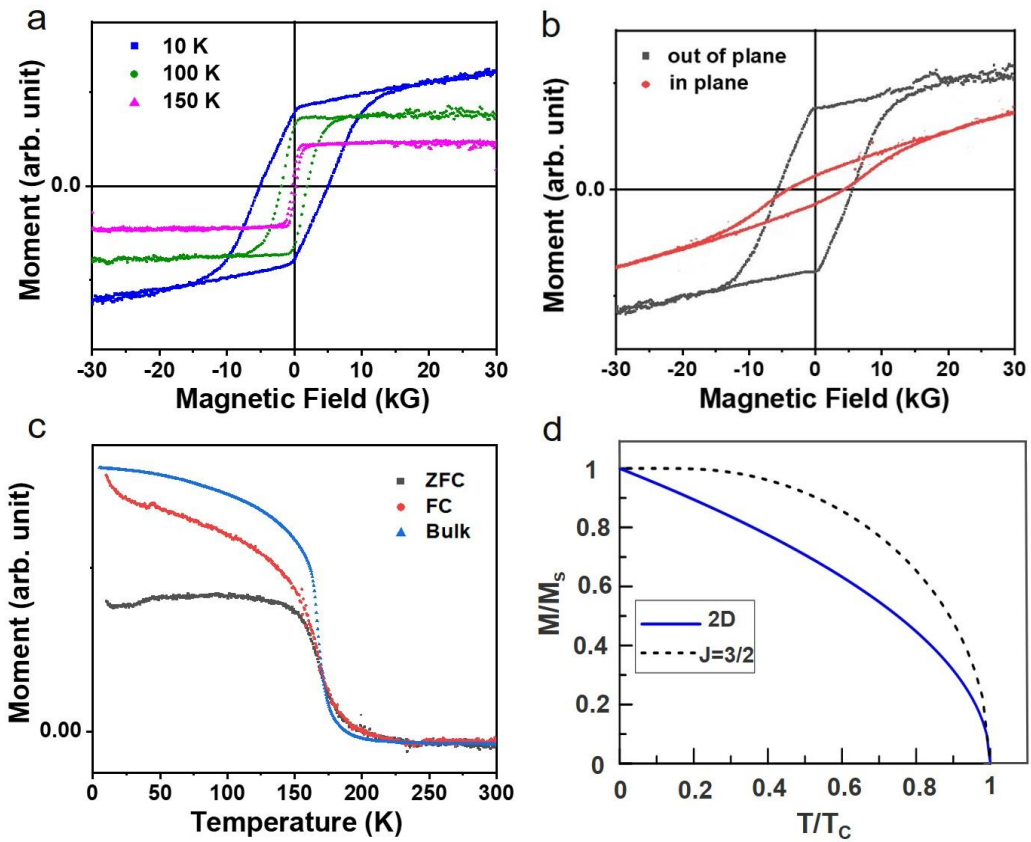

**Figure 4. a,** Out-of-plane magnetic hysteresis loops measured at different temperatures. **b,** Out-of-plane and in-plane magnetic hysteresis loops acquired at 10 K. **c,** ZFC and FC magnetization curves. The FC magnetization of a bulk film is also included for comparison. **d,** Calculated reduced magnetization as a function of reduced temperature of 2D (blue) compared to its 3D counterpart (J=3/2, black) in the mean field approximation.

Figure 4c displays the field-cooled (FC) and zero-field-cooled (ZFC) magnetization as a function of temperature, which confirms the magnetic transition at around 180 K. Due to the small interlayer exchange, the magnetic ordering in the bulk is driven by exchange interactions in the CrTe$_2$ layer, and thus $T_C$ barely changes from bulk to atomically-thin layers. Comparing the FC curves of 2D crystals with that of the bulk film, it can be seen that the magnetization drops faster at low temperatures but more gradually near $T_C$. This is confirmed by calculations based on mean-field approximation (Figure 4d) showing the transition near $T_C$ to be more abrupt for the 3D than the 2D case, mimicking experimental results. At temperatures below 30 K, a sharp rise in magnetization is observed in the FC curve. This is attributed to edge spins with weaker exchange coupling than the interior, leading to disorder [27]. The behavior of the ZFC curve, on the other hand, is quite different from that of conventional ferromagnets. It shows a plateau in a broad temperature range, and decreases slightly with further decreasing temperature. This can be understood as the competition between increasing spin canting and decreasing spin disorder with decreasing temperature.

Polar magneto-optical Kerr effect (MOKE) microscopy was used to extract magnetic properties of individual 2D crystals. The same 2D crystals were then imaged by AFM to correlate their hysteresis behaviors with their morphologies. The MOKE hysteresis at different temperatures on a 2D crystal with a thickness of 11.5 nm shown in **Figure 5**a are plotted in Figure 5c. The behavior is consistent with ensemble measurements with similar $H_C$ values. At low temperatures, the single crystal shows a square loop, with near unity remanence ratio $M_r$ (remanent/saturation magnetization) and a sharp transition near $H_C$, resulting from the perpendicular anisotropy. With increasing temperature, the loops become more rounded, with decreasing $M_r$ and $H_C$. At 170 K, the flake becomes nearly paramagnetic with negligible $H_C$. Temperature dependent $H_C$ plotted in Fig. 5d shows a rapid drop of $H_C$ with increasing temperature. In a material with uniaxial anisotropy, MAE scales with $M_S^3$, which explains the sensitive temperature dependence of $H_C$ [28].

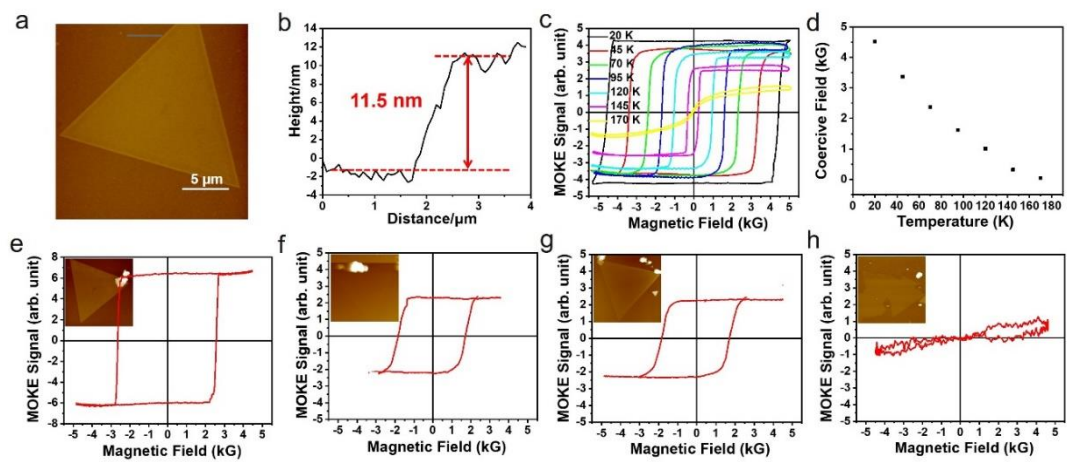

**Figure 5. a,** AFM image and **b,** the corresponding height profile of a single $Cr_2Te_3$ crystal, which shows a thickness of 11.5 nm. The black line in **a** corresponds to the height profile of **b**. **c,** normalized MOKE hysteresis loops and **d,** temperature-dependent coercivity of the corresponding $Cr_2Te_3$ crystal. **e, f, g, h,** MOKE hysteresis loops of $Cr_2Te_3$ 2D crystals with different thicknesses, 25.2 nm, 12.2 nm, 7.8 nm, and 2.8 nm, respectively. e-g were measured at 120 K; h was measured at 20 K. Inset: AFM images of the corresponding $Cr_2Te_3$ crystals.

Hysteresis loops of representative 2D crystals with different thicknesses were further measured, as shown in Figure 5e-h. As the thickness of the flakes decreases from 25.2 to 7.8 nm, the square hysteresis behavior measured at 120 K changes only slightly, owing to the robust perpendicular anisotropy. A small drop in $H_C$ is observed with decreasing thickness. However, a change of behavior is observed for the 2.8 nm crystal (two unit cell thickness). The MOKE signal becomes so weak at 120 K that no clear hysteresis can be detected. At 20 K, while the low signal-to-noise ratio prevents the exact shape of the hysteresis to be resolved, it clearly deviates from the square loops observed for thicker crystals with negligible $M_r$ and $H_C$. This seems to be consistent with the theoretical prediction of a transition from the perpendicular to in-plane anisotropy in the quad-layer $Cr_2Te_3$.

## 3. Conclusion and outlook

In conclusion, 2D ferromagnetic $Cr_2Te_3$ crystals were realized by controlling the kinetic growth in a CVD process. The resulting 2D crystals demonstrate large magnetic anisotropy, which helps to stabilize the ferromagnetic order in 2D $Cr_2Te_3$. First-principles calculations further predict an unconventional half-metallicity, existing only in 2D layers and strengthened by reduced dimensions. Our work represents an important step towards the applications of 2D $Cr_2Te_3$ and related compounds for spintronics, and will inspire the search for new covalent 2D magnets.

**Acknowledgements**

M. B., A. N. K. and M. H. contributed equally to this work. Work supported by US NSF (MRI-1229208, CBET-1510121) and UB VPRED seed grant. Work at the NHMFL supported by NSF DMR-1644779, the State of Florida, and DOE. Y. H. and M. B. thank National Key R&D Program of China (2017YFA0206301), the NSFC (51631001) and the China Postdoctoral Science Foundation (2020M670042). J.L. and M.H. thank the support from NSFC (Grant No.11974156), Guangdong Province (Grant No. 2019A050510001, 2019ZT08C044), Shenzhen Municipality (No. ZDSYS20190902092905285, KQTD20190929173815000), and SUSTech Core Research Facilities and Pico Center.

**Disclosure statement**

No potential conflict of interest was reported by the author(s).